\documentclass[doublecol]{epl2} 

\begin{document}

\title{Anyonic $\mathcal{PT}$ symmetry, drifting potentials and non-Hermitian delocalization}
\shorttitle{ Anyonic $\mathcal{PT}$ symmetry ... } 

\author{S. Longhi \inst{1,2} \and E. Pinotti \inst{1}}
\shortauthor{S. Longhi}

\institute{                    
  \inst{1}  Dipartimento di Fisica, Politecnico di Milano, Piazza L. da Vinci 32, I-20133 Milano, Italy\\
  \inst{2}  Istituto di Fotonica e Nanotecnlogie del Consiglio Nazionale delle Ricerche, sezione di Milano, Piazza L. da Vinci 32, I-20133 Milano, Italy\\
   }

\abstract{We consider wave dynamics for a Schr\"odinger equation with a non-Hermitian Hamiltonian $\mathcal{H}$ satisfying the generalized (anyonic) parity-time symmetry $\mathcal{PT H}= \exp(2 i \varphi) \mathcal{HPT}$, where $\mathcal{P}$ and $ \mathcal{T}$ are the parity and time-reversal operators.
  For a stationary potential, the anyonic phase $\varphi$ just rotates the energy spectrum of $\mathcal{H}$ in complex plane, however for a drifting potential the energy spectrum is deformed and the scattering and localization properties of the potential show intriguing behaviors arising from the breakdown of the Galilean invariance when $\varphi \neq 0$. In particular, in the unbroken $\mathcal{PT}$ phase the drift makes a scattering potential barrier reflectionless, whereas for a potential well the number of bound states decreases as the drift velocity increases because of a non-Hermitian  delocalization transition.} 
\maketitle

\section{Introduction}
The concept of parity-time ($\mathcal{PT}$) symmetry, originally introduced by Carl Bender in non-Hermitian extensions of conventional quantum mechanics \cite{r1,r2,r3}, is attracting a great attention in different areas of physics \cite{r4,r6,r7,r8,r12}. The Hamiltonian $\mathcal{H}$ is $\mathcal{PT}$ symmetric if $\mathcal{HPT}= \mathcal{PTH}$, where $\mathcal{P}$ and $\mathcal{T}$ are the parity and time reversal operators. Like for the broader class of pseudo-Hermitian Hamiltonians \cite{r9,r10,r11}, $\mathcal{PT}$ symmetric Hamiltonians possess an entirely real energy spectrum below a symmetry breaking phase transition.  Besides $\mathcal{PT}$ symmetry, anti-$\mathcal{PT}$ (or Wick-rotated $\mathcal{PT}$) symmetry, i.e. $\mathcal{HPT}= -\mathcal{PTH}$, as well as other symmetries, such as $\mathcal{CPT}$, have been introduced and demonstrated in some recent works \cite{r13,r14,r15,r16,r17,r18,r19,r19bis}. Owing to the non-uniqueness of parity and time reversal operators, different symmetries can coexist  in the same system \cite{r19bis,refB,long}. $\mathcal{PT}$ symmetry and antisymmetry can be regarded as special cases of the generalized $\mathcal{PT}$ symmetry $\mathcal{HPT}= \exp(2i \varphi) \mathcal{PTH}$ when the phase $\varphi$ takes the value $0$ and $\pi/2$, respectively. The commutation relation $\mathcal{HPT}= \exp(2i \varphi) \mathcal{PTH}$ is formally similar to the one for creation/destruction operators of quasi-particles (anyons) with a statistics intermediate between bosons ($\varphi=0$) and fermions ($\varphi= \pi/2$). Therefore, the generalized 
$\mathcal{PT}$ symmetry $\mathcal{HPT}= \exp(2i \varphi) \mathcal{PTH}$ can be referred to as {\em anyonic} $\mathcal{PT}$ symmetry, although we are not actually dealing with anyonic quasi-particles. For a stationary Hamiltonian of Schr\"odinger type, anyonic $\mathcal{PT}$ symmetry is rather generally obtained by multiplying a $\mathcal{PT}$-symmetric Hamiltonian by a phase factor, which rotates in complex plane the energy spectrum. Such a rotation may have major physical implications, for example it can provide symmetry protection of non-Hermitian zero-energy modes \cite{refB}. Here we disclose  intriguing scattering and localization effects in anyonic $\mathcal{PT}$ symmetric Hamiltonians  that arise when considering a {\em drifting potential}: the absence of reflection for a moving potential barrier and the disappearance of bound states for a fast moving potential well. Such phenomena can be traced back to breakdown of the Galilean invariance of the Schr\"odinger equation for a non vanishing anyonic phase and to the phenomenon of non-Hermitian delocalization \cite{r20,r21,r22,r23,r24}.
\section{Anyonic parity-time symmetry} 
Let us consider wave dynamics described by a Schr\"odinger-like equation with a non-Hermitian and time-dependent Hamiltonian $\mathcal{H}(T)$ for the wave function $\psi(X,T)$
\begin{equation}
i \frac{\partial \psi}{\partial T}= \mathcal{H}(T) \psi
\end{equation}
defined in the $L^2(R)$ Hilbert space. 
The Hamiltonian $\mathcal{H}$ is said to satisfy a generalized (anyonic) $\mathcal{PT}$ symmetry if 
\begin{equation}
\mathcal{PTH}=\exp(2i \varphi) \mathcal{HPT}
\end{equation}
for some angle $\varphi$, with $0 \leq \varphi \leq \pi/2$, where $\mathcal{T}$ and $\mathcal{P}$ are the time-reversal and parity operators, defined in the usual way as \cite{r3,r4} $\mathcal{T} \psi(X,T)=\psi^*(X,-T)$ and $\mathcal{P} \psi(X,T)=\psi(-X,T)$. Note that the ordinary $\mathcal{PT}$ symmetry \cite{r3} is retrieved when $\varphi=0$, whereas for $\varphi= \pi/2$ one has anti-$\mathcal{PT}$ symmetry. A stationary or time-dependent Hamiltonian $\mathcal{H}=\mathcal{H}_0$ of the form
\begin{equation}
\mathcal{H}_0=-\frac{\partial^2}{\partial X^2}+V(X-vT),
\end{equation}
where $v$ is a drift velocity, satisfies the ordinary $\mathcal{PT}$ symmetry [Eq.(2) with $\varphi=0$] provided that
\begin{equation}
V(-X)=V^*(X)
\end{equation}
for the complex potential $V(X)$. If $\mathcal{H}_0$ is Hermitian or satisfies the ordinary $\mathcal{PT}$ symmetry, $\mathcal{PTH}_0=\mathcal{H}_0 \mathcal{P T}$, then the phase-rotated Hamiltonian $\mathcal{H}=\exp(-i \varphi) \mathcal{H}_0$, i.e.
 \begin{equation}
 \mathcal{H}= -\exp(-i \varphi) \frac{\partial^2}{\partial X^2}+\exp(-i \varphi) V(X-vT)
  \end{equation}
  satisfies the anyonic $\mathcal{PT}$ symmetry [Eq.(2)] with anyonic phase $\varphi$. 
In the following, we will assume that $V(X)$ is a short-range potential with $V(X) \rightarrow 0$ as $X \rightarrow \pm \infty$, and that $\mathcal{H}_0$ is either Hermitian or in the unbroken $\mathcal{PT}$ phase, i.e. the energy spectrum of $\mathcal{H}_0$ (assuming a stationary potential $v=0$) is entirely real. The energy spectrum of $\mathcal{H}_0$ comprises rather generally a point spectrum corresponding to bound states of negative energies, and a continuous spectrum of scattered states with positive energies [Fig.1(a)]. Clearly, for a stationary potential $v=0$ the energy spectrum of the phase-rotated Hamiltonian $\mathcal{H}$, defined by Eq.(5), is rotated clockwise by an angle $\varphi$ in complex plane as compared to the spectrum of $\mathcal{H}_0$, as shown in Fig.1(b). 
Note that, if the potential $V_0(X)$ is real on the real $X$ line with $V_0(-X)=V_0(X)$ and the analytic continuation $V(Z)$ of $V_0$ in the $Z=X+iY$ complex plane is holomorphic in the stripe $|Y|<\Delta$, then an anyonic $\mathcal{PT}$-symmetric Hamiltonian $\mathcal{H}$ in the unbroken $\mathcal{PT}$ phase can be synthesized according to Eq.(5) assuming the potential 
\begin{equation}
V(X)=V(Z=X-i \delta), 
\end{equation}
where $\delta$ is an arbitrary shift constrained by $|\delta|< \Delta$. In fact, the imaginary spatial displacement $i \delta$ inside the domain of analyticity of $V$ does not change the energy spectrum of $\mathcal{H}_0$ with respect to the Hermitian limit $\delta=0$. 
\begin{figure}
\onefigure[width=7.8cm]{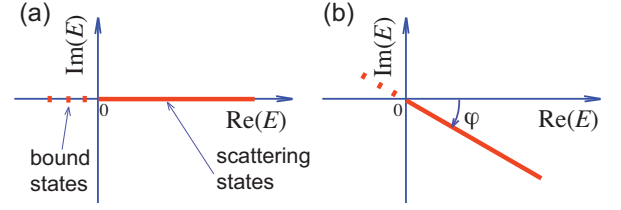}
  \caption{ (a) Energy spectrum of a stationary $\mathcal{PT}$-symmetric Hamiltonian with $\varphi=0$ in the unbroken $\mathcal{PT}$ phase for a complex potential $V(-X)=V^*(X)$ that vanishes as $X \rightarrow \pm \infty$. 
   The energy spectrum comprises rather generally the point spectrum of bound states (isolated circles, negative energies) and the absolutely continuous spectrum of scattering states (positive energies, solid curve). (b) Energy spectrum of a stationary $\mathcal{PT}$-symmetric Hamiltonian with anyonic phase $\varphi \neq 0$ in the unbroken $\mathcal{PT}$ phase. A non-vanishing anyonic phase $\varphi$ just rotates the energy spectrum around the origin of the complex energy plane.}
\end{figure}
 \section{Drifting potentials and Galilean invariance}
  Let us consider the Schr\"odinger equation described by the Hamiltonian (5) having an arbitrary  potential $V(X)$ -- not necessarily satisfying the $\mathcal{PT}$ condtion (4)-- drifting in time at some speed $v$. For $\varphi=0$, the drift does not introduce any appreciable effect on wave dynamics owing to the Galilean invariance of the Schr\"odinger equation \cite{r26,r27}. However, for $\varphi \neq 0$ Galilean invariance of the Schr\"odinger equation is broken, as shown below.  This implies that in the anyonic phase a drifting potential causes a deformation of the energy spectrum of the Hamiltonian in the moving reference frame and deeply changes the scattering properties of the potential. To prove the above statements, let us consider the wave dynamics in the moving reference frame $(x,t)$ defined by the Galilean boost
\begin{equation} 
x=X-vT \; , \;\; t=T.
\end{equation}
In the $(x,t)$ frame, the Schr\"odinger equation (1) reads
\begin{equation}
i \frac{\partial \psi}{\partial t}= \mathcal{H}_{eff} \psi
\end{equation}
for the wave function $\psi=\psi(x,t)$, where the stationary Hamiltonian $\mathcal{H}_{eff}$ is given by
\begin{equation}
\mathcal{H}_{eff}=-\exp(-i \varphi) \frac{\partial^2}{\partial x^2}+ \exp(-i \varphi) V(x)+iv \frac{\partial}{\partial x}.
\end{equation}
To obtain invariance of the Schr\"odinger equation for the Galilean boost (7), we need  to remove the drift term $i v( \partial / \partial x)$ appearing in $\mathcal{H}_{eff}$. This can be {\em formally} achieved by introduction of the (rather generally {\em non-Hermitian}) gauge transformation
\begin{equation}
\psi(x,t)= \phi(x,t) \exp(i \alpha x-i \beta t)
\end{equation}
where we have set
\begin{equation}
\alpha \equiv \frac{v}{2} \exp(i \varphi) \; , \;\; \beta \equiv -\frac{v^2}{4} \exp(i \varphi).
\end{equation}
In fact, substitution of the Ansatz (10) into Eq.(8) and using Eqs.(9) and (11) yields
\begin{equation}
i \frac{\partial \phi}{\partial t}= -\exp(-i \varphi) \frac{\partial^2 \phi}{\partial x^2}+ \exp(-i \varphi) V(x) \phi= \exp(-i \varphi) \mathcal{H}_0 \phi
\end{equation}
which is precisely the Schr\"odinger equation with a stationary potential. For $\varphi=0$, the gauge transformation (10) does not modify the Hilbert space of wave functions and we retrieve the well-known Galilean invariance of the Schr\"odinger equation \cite{r26,r27}.
However, for a non vanishing phase $\varphi$ the gauge transformation (10) introduces an unbounded time-dependent operator, namely
\begin{equation}
|\psi(x,t)|^2=|\phi(x,t)|^2 \exp(-v x \sin \varphi ) \exp \left[ \frac{v^2}{2}  t \sin \varphi \right].
\end{equation}
Since the Schr\"odinger operator $\mathcal{H}_{eff}$ should be defined on the $L^2(R)$ Hilbert space, for $\varphi \neq 0$ the gauge transformation (10) cannot be used and thus the Galilean invariance of the Schr\"odinger equation is broken. Note that such a result, i.e. breakdown of the Galilean invariance for $\varphi \neq 0$, is a rather general one and holds for arbitrary potential shapes $V(X)$, i.e.not necessarily satisfying the $\mathcal{PT}$ condition (4).

 \section{Scattering and non-Hermitian delocalization}
Wave scattering off from non-Hermitian potentials is known to show distinctive features as compared to Hermitian potentials, such as asymmetric reflection for left/right incidence sides  \cite{r27bis,r27tris}. Here we disclose new intriguing behaviors arising from potential drift and rooted in the breakdown of the Galilean invariance discussed above. While the results derived in this section are rather general ones, in the following we focus our analysis to the case of either $\mathcal{H}_0$ Hermitiann (i.e. $V(X)$ real)  or  $\mathcal{H}_0$  $\mathcal{PT}$ symmetric in the unbroken $\mathcal{PT}$ phase. This ensures that, at $\varphi=0$, the energy spectrum of $\mathcal{H}= \mathcal{H}_0$ is entirely real.\\
Let us indicate by $u_1(X)$,$u_2(X)$, .., $u_N(X)$ the $N$ bound states with real and negative energies $E_1<E_2< ...<E_N<0$ sustained by the $\mathcal{PT}$ symmetric Hamiltonian $\mathcal{H}_0$ with stationary potential, defined by Eq.(3) with $v=0$ and $V(-X)=V^*(X)$, and by $E(k)=k^2$ the continuous spectrum of scattering states (asymptotic plane waves) with wave number $k$. If the potential $V(X)$ does not sustain  
bound states, such as in case of a potential barrier, $N=0$ and the spectrum is entirely continuous. For the phase-rotated anyonic $\mathcal{PT}$-symmetric Hamiltonian $\mathcal{H}=\exp(-i \varphi) \mathcal{H}_0$, in the absence of drifts ($v=0$) the energy spectrum is rotated clockwise by the anyonic angle $\varphi$ [Fig.1(b)], but it is not deformed. In particular, a non-vanishing anyonic phase does not change the number of bound states of the Hamiltonian, despite the ground-state $u_1(x)$ becomes the dominant mode, i.e. the mode with the largest imaginary part of the energy. Let us now consider the case of a drifting potential. In the moving reference frame $(x,t)$, the wave function $\psi(x,t)$ satisfies the Schr\"odinger equation (8) with the Hamiltonian $\mathcal{H}_{eff}$ given by Eq.(9). Owing to breakdown of Galilean invariance, for $\varphi \neq 0$ the drift term $i v \partial / \partial x$ in Eq.(9) can not be removed; as a consequence, the spectrum of $\mathcal{H}_{eff}$ is not obtained from the one of $\mathcal{H}_{0}$ by a mere phase rotation in complex plane. How does the spectrum of $\mathcal{H}_{eff}$ look like for $\varphi \neq 0$? \par
(i) {\em Continuous spectrum.} Since $V(x) \rightarrow 0$ as $x \rightarrow \pm \infty$, from Eq.(9) it readily follows that the scattering states of $\mathcal{H}_{eff}$ are asymptotically plane waves 
with wave number $k$ and energy $\tilde{E}(k)$ given by the dispersion curve
\begin{equation}
\tilde{E}(k)=\exp(-i \varphi) k^2-kv.
\end{equation}
Note that the continuous spectrum $\tilde{E}(k)$ of $\mathcal{H}_{eff}$ is not merely obtained by the rotation in complex plane, by the angle $\varphi$, of the spectrum $E(k)=k^2$ of $\mathcal{H}_0$ as in Fig.1(b), because of the additional contribution arising from the drift term. As a result, the continuous spectrum of $\mathcal{H}_{eff}$ in complex plane is not a straight line anymore, but becomes a bent curve, as shown in Fig.2. \par
(ii) {\em Point spectrum.}  In view of the transformation defined by Eqs.(10) and (11), it is clear that $\tilde{u}_n(x) \equiv u_n(x) \exp(i\alpha x)$ is formally a solution to the equation
\begin{equation}
\mathcal{H}_{eff}\tilde{u}_n(x)=\tilde{E}_n \tilde{u}_n(x)
\end{equation} 
with 
\begin{equation}
\tilde{E}_n=E_n \exp(-i \varphi)+\beta
\end{equation}
($n=1,2,...,N$). However, this does not necessarily imply that $\tilde{E}_n$ belongs to the point spectrum of $\mathcal{H}_{eff}$. In fact, owing to the exponential term introduced by the transformation [Eq.(13)], the wave function $\tilde{u}_n(x)$ can become delocalized (non-normalizable) for a drift velocity $|v|$ larger than a critical value $v_c$, i.e. the bound state can disappear for a fast enough drift of the potential. This effect is analogous to the non-Hermitian delocalization transition found in non-Hermitian Anderson models or for quantum particles in a potential well driven by an imaginary magnetic field \cite{r20,r21,r22,r23,r24,r28,r29}. To determine the critical drift velocity $v_c$, above which the bound state disappears,  let us notice that, as $x \rightarrow \pm \infty$, the bound state $u_n(x)$ has the asymptotic decay behavior $u_n(x) \sim \exp(- \sqrt{|E_n|} |x|)$. Therefore, the critical drift velocity $v_c$ is obtained by letting $\sqrt{|E_n|}=|(v/2) \sin \varphi |$, i.e.
\begin{equation}
v_c= 2 \sqrt{|E_n|}/ \sin \varphi .
\end{equation}
\begin{figure}
\onefigure[width=7.8cm]{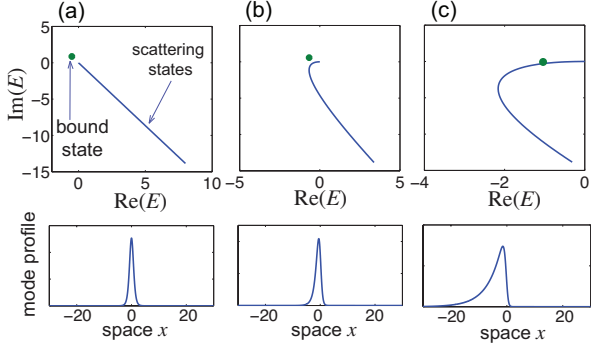}
  \caption{
    Upper panels: Energy spectrum of the $\mathcal{PT}$-symmetric P\"oschl-Teller potential well $V(x)=-2/ \cosh^2(x-i \delta)$ for anyonic phase $\varphi=\pi/3$ and for increasing values of the drift velocity $v$: (a) $v=0$, (b) $v=0.5 v_c$ and (c) $v=0.9 v_c$, where $v_c=2/ \sin \varphi$ is the critical drift velocity. The solid curve is the absolutely continuous spectrum of scattering states, whereas the circle is the energy of the bound state. Lower panels: profiles of the bound state $|\tilde{u}_1(x)|^2$ sustained by the P\"oschl-Teller potential well  for $\delta=0.2$ and for the three above values of the drift velocities. As $v$ approaches $v_c$, the bound state delocalizes and its energy coalesces with the continuous spectrum.}
\end{figure}
As the drift velocity $v$ approaches $v_c$ from below, the bound state $\tilde{u}_n(x)$ becomes less and less confined in one spatial direction, and its corresponding eigenvalue $\tilde{E}_n$ coalesces with the continuous spectrum. In fact, from Eqs.(11) , (14) and (16) it can be readily shown that, at $v=v_c$, one has $\tilde{E}_n=\tilde{E}(k_c)$ for the  critical wave number $k_c=\sqrt{-E_n} / \tan \varphi$. This means that, as the drift velocity is increased, the number of bound states can be reduced and fully cancelled. This is illustrated, as an example, in Fig.2 for the P\"oschl-Teller potential well \cite{r30}
\begin{equation}
V(x)=-\frac{\nu (\nu+1)}{\cosh^2(x-i \delta)}
\end{equation}
($\nu >0$). 
As is well-known, this potential is $\mathcal{PT}$ symmetric in the unbroken $\mathcal{PT}$ phase for $|\delta| < \pi/2$, and sustains $N=1+[ \nu ]$ bound states with energies $E_n=-(\nu-n+1)^2$ ($n=1,2,...,N$), where $[ \nu ]$ is the integer part of $\nu$. For $\nu$ integer, the potential is reflectionless and there is just one unbound state with zero energy $E=E_N=0$. Figure 2 shows, as an example, non-Hermitian delocalization and coalescence of the point spectrum in the continuum as the drift velocity is increased to approach the critical value $v_c$ for the P\"oschl-Teller potential well with $\nu=1$, which sustains one bound state. We note that a change of number of bound states in Hamiltonian models with drifting potentials and arising from breakdown of the Galilean invariance can be observed in Hermitian models as well, such as in discrete wave mechanics \cite{r31}, however in the present work the disappearance of bound states for increasing drift velocities is ultimately ascribed to the phenomenon of non-Hermitian delocalization, i.e. it is a clear signature of non-Hermitian dynamics. \par

From the dynamical viewpoint, the drift of the potential in the anyonic phase gives rise to some interesting effects, such as non-Hermitian transparency of a potential barrier and instability of the dominant bound state in a potential well near the critical drift velocity in spite of eigenvalue stability. Let us discuss in details such two effects.\par
\begin{figure}
  \onefigure[width=7.8cm]{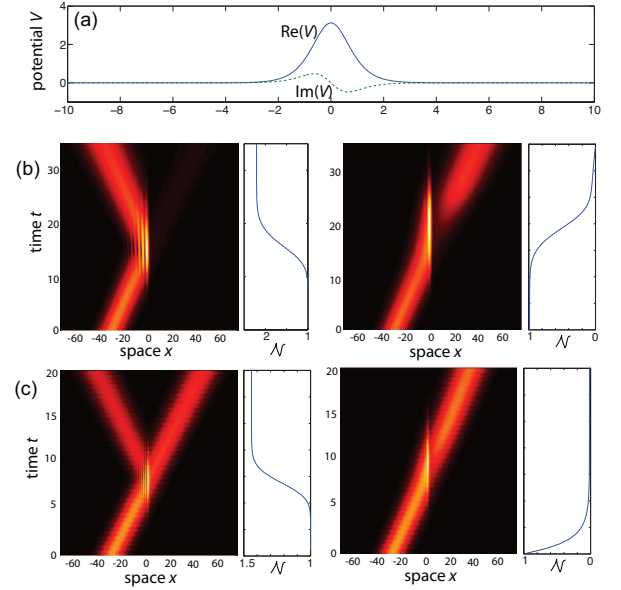}
\caption{ 
    Scattering of a Gaussian wave packet from the complex potential barrier $V(x)=V_0/ \cosh^2(x-i \delta)$ for parameter values $V_0=3$, $\delta=-0.5$ and drift velocity $v=-2$. Panel (a) shows the profile of the scattering barrier (real and imaginary parts of the potential). Panels (b) and (c) show on a pseucocolor map the temporal evolution of the normalized wave packet distribution $| \psi(x,t)|^2 / \mathcal{N}$ and of the norm $\mathcal{N}(t) \equiv \int dx |\psi(x,t)|^2$ for the initial condition $\psi(x,0) \propto \exp [-(x-d)^2 /w^2+ikx ] $ ($w=10$, $d=32$) with carrier wave number $k=0$ in (b), and $k=1$ in (c). In (b) and (c), left and right panels refer to the different anyonic phase $\varphi=0$ and $\varphi= \pi/8$, respectively. In the former case the barrier reflects, partially or totally, the incident wave packet, whereas in the latter case the barrier is reflectionless.}
\end{figure}
{\it Non-Hermitian barrier transparency.} Let us consider a potential barrier $V=V(x)$ vanishing at $x \rightarrow \pm \infty$ and let us consider, in the moving reference frame $(x,t)$, a plane wave with wave number $k$ and group velocity $v_g= {\rm Re} (d \tilde{E}/dk)=2 k \cos \varphi-v$ coming from $x=-\infty$ and incident onto the barrier from the left side. The barrier drifts at the speed $v$. We typically assume $v<0$, so that $v_g>0$ for any $k \geq 0$ and anyonic phase $\varphi$. As in usual scattering problems, a solution to Eq.(8) that describes the scattering of the incident plane wave has the asymptotic form
\begin{equation}
\psi(x,t) \sim \left\{
\begin{array}{l}
\left[ \exp(ikx)+r(k) \exp(i k_r x) \right] \exp[-i \tilde{E}(k)t]   \\
 \;\;\;\;\;\;\;\; \;\;\;\;\;\;\;\; \;\;\;\;\;\;\;\; \;\;\;\;\;\;\;\;   x \rightarrow -\infty  \\
t(k) \exp [ ikx-i \tilde{E}(k)t ]  \;\;\;\;\;\;\;\; \;\;\;\;\;\;\;\;   x \rightarrow \infty 
\end{array}
\right.
\end{equation}
where $r=r(k)$ and $t=t(k)$ are the spectral reflection and transmission coefficients, respectively, $\tilde{E}(k)$ is given by Eq.(14), and $k_r$ is the wave number of the reflected wave, which is determined by the elastic scattering condition $\tilde{E}(k_r)=\tilde{E}(k)$ with ${\rm Re} (d \tilde{E}/dk)_{k_r}<0$, namely
\begin{equation}
k_r=-k+v \exp(i \varphi).
\end{equation}
 Clearly, for conventional $\mathcal{PT}$-symmetry ($\varphi=0$), the wave number $k_r$ turns out to be real and the reflected wave is propagative with a group velocity opposite to the one of the incident wave. In the laboratory reference frame $(X,T)$ this leads to the usual Doppler shift of frequency of reflected wave from a moving potential barrier. Unless for special tailoring of $V(x)$, the potential barrier does reflect some of incident wave. However, for a non vanishing anyonic phase $\varphi \neq 0$, from Eq.(20) it follows that the wave number $k_r$ of reflected wave has a nonvanishing imaginary part, indicating that the reflected wave is {\it evanescent}.  This means that, on the far left from the scattering potential, the reflected wave decays exponentially in space and the potential barrier effectively becomes {\it reflectionless}. Remarkably, such a reflectionless property is independent of the potential barrier shape, and it is thus rather distinct than the reflectionless property showed by some special complex potentials studied in the recent literature, such as potentials with spatial Kramers-Kronig profiles \cite{r32,r33}. While in the latter case waves are not reflected because the scattering potential shows a one-sided spatial Fourier spectrum and is unable to create regressive waves, in our case waves are not reflected because of a {\it non-Hermitian transparency effect}, quite similar to the one found in tight-binding lattices in the presence of an imaginary gauge field \cite{r29}: reflected waves are evanescent ones. An example of reflectionless drifting potential barrier induced by a non vanishing anyonic phase is shown in Fig.3. The figure illustrates the scattering of a Gaussian wave packet off from the potential barrier $V(x)=V_0/ \cosh^2(x-i \delta)$ ($V_0>0$, $|\delta|<\pi/2$) for two different carrier wave numbers $k$ for conventional $\mathcal{PT}$ symmetry ($\varphi=0$, left panels) and for anyonic $\mathcal{PT}$ symmetry ($\varphi=\pi/8$, right panels). Clearly, while in the former case the wave packet is partially or almost totally reflected from the barrier, in the latter case the barrier is reflectionless. \par
 {\it Anomalous amplification of perturbations and instability of the dominant bound state.} Let us consider a potential well sustaining (for the sake of simplicity) one bound state, such as the P\"oschl-Teller potential well (18) with $\nu \leq 1$ and $|\delta| < \pi/2$. Clearly, for a drift velocity $v$ below the critical value $v_c$, the imaginary part of the energy $\tilde{E}_1$ of the bound state $\tilde{u}_1(x)=u_1(x) \exp(i \alpha x)$ remains larger than the one of scattered states $\tilde{E}(k)$, i.e. the bound state is the dominant and linearly (eigenvalue) stable mode. However, since the non-Hermitian Hamiltonian $\mathcal{H}_{eff}$ can show non-normal dynamics \cite{r16,r18,r34}, the spectral dominance of the bound state and the energy spectrum of $\mathcal{H}_{eff}$ may have little to do with the dynamical behavior of the system, which is at best measured by the norm  \cite{r34} 
 \begin{equation}
  G_t \equiv \| \exp[ -i (\mathcal{H}_{eff}-\tilde{E}_1) t] \|^2.
  \end{equation}
  $G_t$ measures the largest power amplification, at a given time $t$, that an initial perturbation can undergo, with respect to the bound state power level.
For a normal operator, $G_t$ can never exceed one, however for a highly non-normal operator $G_t$ can take large values, driving the system into a kind of unstable behavior in spite of the eigenvalue stability of the bound state. This behavior, induced by non-normal dynamics, i.e. by non-orthogonality of operator eigenfunctions, is known to arise for hydrodynamic flows \cite{r37} and laser systems \cite{r38,r39}, where huge values of $G_t$ can even  drive the system into turbulence. A simple analytical expression of the power growth $G_t$ can be obtained in the $t \rightarrow \infty$ limit. In this limit, $G_{\infty}$ equals the so-called Petermann excess noise factor and the perturbation that undergoes the largest amplification is the bound state  $u_1^{\prime \dag}(x)$ of the adjoint Hamiltonian 
\begin{equation}
\mathcal{H}_{eff}^{\dag}=-\exp(i \varphi) \frac{\partial^2}{\partial x^2}+ \exp( i \varphi) V(-x)+iv \frac{\partial}{\partial x}.
\end{equation}
The value of $G_{\infty}$ can be readily calculated and reads 
\begin{equation}
G_{\infty}={\langle \tilde{u}_1 | \tilde{u}_1   \rangle \langle \tilde{u}^{\dag}_1 | \tilde{u}^{\dag}_1  \rangle} /  {\left|  \langle \tilde{u}^{ \dag}_1 | \tilde{u}_1 \rangle  \right|^2}
\end{equation}
where $\langle f | g \rangle \equiv \int dx f^*(x) g(x)$ is the usual scalar product in $L^2$. Taking into account that in our case
\begin{eqnarray}
\tilde{u}_1(x) & = & u_1(x) \exp [i (v x/2) \exp(i \varphi)] \\
\tilde{u}_1^{\dag}(x) & = & u_1^*(x) \exp [i (vx /2) \exp(-i \varphi)] 
\end{eqnarray}
 substitution of Eqs.(25) and (26) into Eq.(24) finally yields
\begin{eqnarray}
G_{\infty} & =& \int_{-\infty}^{\infty}dx |u_1(x)|^2 \exp (-v x \sin \varphi) \nonumber \\
& \times & \frac{ \int_{-\infty}^{\infty}dx |u_1(x)|^2 \exp (v x \sin \varphi)}{ \left| \int_{-\infty}^{\infty}dx u_1^2(x) \right|^2}. 
\end{eqnarray}
In the above equations, $u_1(x)$ is the bound state of the $\mathcal{PT}$-symmetric Hamiltonian $\mathcal{H}_0$, i.e. 
\begin{equation}
-\frac{d^2u_1}{dx^2}+V(x) u_1(x)=E_1 u_1(x).
\end{equation}
Equation (26) clearly shows that large values of $G_{\infty}$, i.e. large power amplification of perturbations that could destabilize the dominant bound state, can be obtained in two cases: (i) when the 
$\mathcal{PT}$ symmetry is close to the symmetry breaking transition, signaled by the appearance of an exceptional point and the self-orthogonality of the eigenmode $ \int_{-\infty}^{\infty} dx u_1^2(x)=0$; or
(ii) when the drift velocity $v$ approaches the critical value $v_c$. In the former case the bound state remains localized and $G_{\infty}$ becomes large near the exceptional point because of the self-orthogonality 
 of the mode. This happens even for a stationary (i.e. not drifting) potential and does not require anyonic $\mathcal{PT}$ symmetry (see, for example, \cite{r41}). In the latter case a large value of $G_{\infty}$ arises because of the non-Hermitian delocalization of the bound state as $v \rightarrow v_c$, i.e. this kind of large transient amplification strictly requires a drifting potential and a nonvanishing anyonic phase $\varphi$ of the $\mathcal{PT}$ symmetry.
 \begin{figure}[htb]
\onefigure[width=7.8cm]{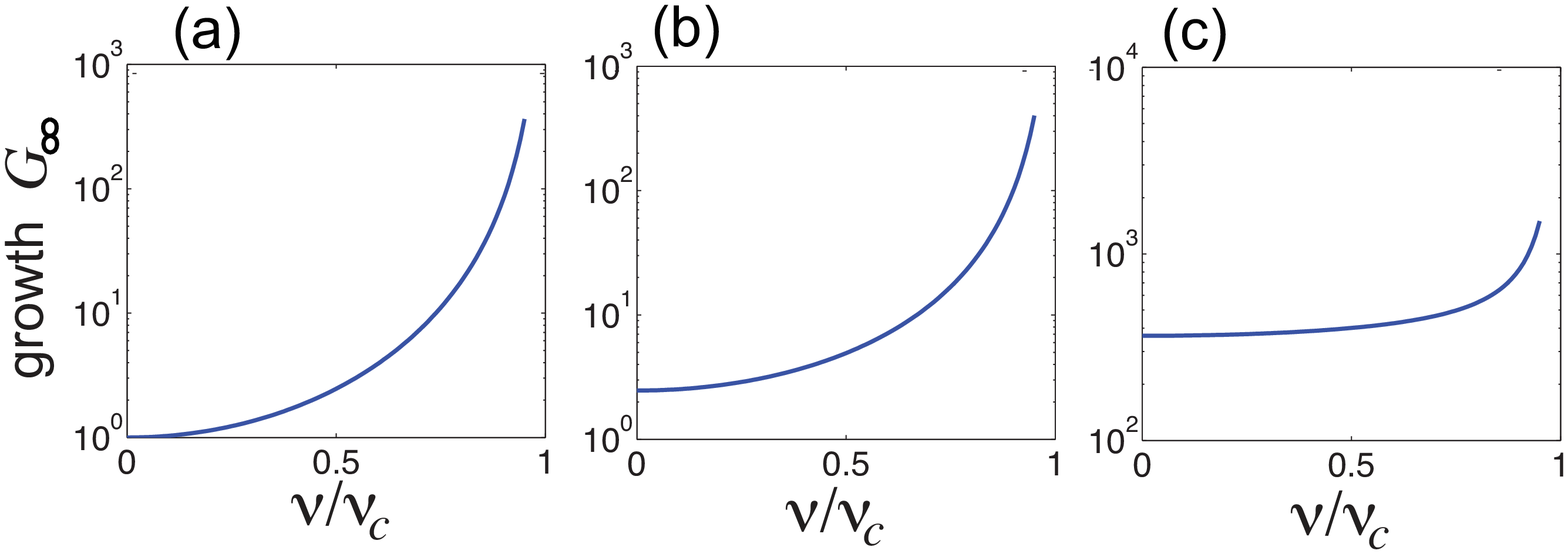}
\caption{ Behavior of the growth (amplification) factor $G_{\infty}$ of perturbations versus the drift velocity $v$ for the P\"oschl-Teller potential well $V(x)=-2/ \cosh^2(x-i \delta)$ for anyonic phase $\varphi= \pi/3$ and for a few increasing values of $\delta$: (a) $\delta=0$ (Hermitian limit), (b) $\delta= \pi/4$ and (c) $\delta=0.9 \times \pi/2$. The critical drift velocity $v_c$ is $v_c=2/\sin \varphi=4 / \sqrt{3}$.}
\end{figure} 
 \begin{figure}[htb]
\onefigure[width=7.8cm]{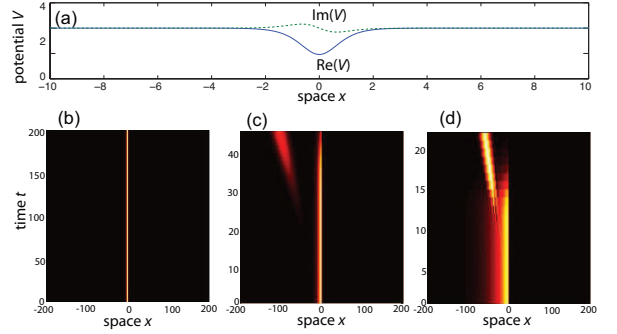}
 \caption{ Non-normal dynamics and transient amplification of perturbations in a drifting P\"oschl-Teller potential well [Eq.(18) with $\nu=1$ and $\delta=0.2$], sustaining one bound state, for increasing values of the drift velocity. The anyonic phase is $\varphi= \pi/3$. (a) Profile of the $\mathcal{PT}$ symmetric potential well (real and imaginary parts). (b-d): Numerically-computed evolution of the normalized wave function distribution $|\psi(x,t)|^2 / \int_{-\infty}^{\infty} dx |\psi(x,t)|^2$ for a few increasing values of the drift velocity: $v=0.2 v_c$ in (b), $v=0.8 v_c$ in (c), and $v=0.95 v_c$ in (d), where $v_c=2 / \sin \varphi$ is the critical drift velocity.  The initial condition is the bound state of the potential well.}
\end{figure}  
For example, let us consider the P\"oschl-Teller potential well (18) with $\nu=1$ and $0 \leq \delta < \pi/2$; the bound state $u_1(x)$ can be calculated in a closed form and reads 
\begin{equation}
u_1(x)=1 / {\cosh (x-i \delta)}.
\end{equation}
 $\mathcal{PT}$ symmetry breaking is reached as $\delta \rightarrow \pi/2^-$. In fact, after setting $\delta=\pi/2- \epsilon$ with $\epsilon>0$ a small quantity, in the $\epsilon \rightarrow 0$ limit one has:
 \begin{equation}
 V(x) \simeq 2/ (x+i \epsilon)^2
 \end{equation}
 and 
 \begin{equation}
 u_1(x) \sim 1/(x+i \epsilon)
 \end{equation}
 which is a self-orthogonal mode, corresponding to an exceptional point of $\mathcal{H}_0$ at the edge of the continuum \cite{r41}. Figure 4 shows the behavior of the power growth $G_{\infty}$ for the potential well with $\varphi= \pi/3$ anyonic phase versus the drift velocity $v$ for a few increasing values of the complex space displacement $i \delta$. For $\delta=0$ or far from the exceptional point [Figs.4(a) and (b)], $G_{\infty}$ takes large values only when the drift velocity approaches the critical value $v_c$. On the other hand, when $\delta$ is pushed close to the exceptional point $\delta= \pi/2$ the power amplification $G_{\infty}$ is less sensitive to the drift velocity and can take large values also for a potential well at rest [Fig.4(c)]. In the regime of a large amplification $G_{\infty}$, even if the bound state is the dominant mode with the largest value of imaginary part of energy , small initial perturbations or noise in the system can  effectively destabilize the mode \cite{r37,r38}. 
 In numerical simulations of the Schr\"odinger equation, discretization or truncation effects and numerical accuracy are often enough to observe disruption of the bound mode. As an example, Fig.5
shows the numerically-computed evolution of the normalized wave distribution $| \psi(x,t)|^2 / \int dx |\psi(x,t)|^2$, in the moving $(x,t)$ reference frame, for a drifting P\"oschl-Teller potential well [Eq.(19) with $\nu=1$ and $\delta=0.2$] in the anyonic phase $\varphi=\pi/4$ for a few increasing values of the drift velocity $v$, below the critical velocity $v_c$. The profile of the $\mathcal{PT}$-symmetric potential well (real and imaginary parts)  is shown in Fig.5(a). The system is initially prepared in the exact bound state $\psi(x,0)=\cosh^{-1}(x-i \delta) \exp[i (vx/2) \exp(i \varphi)]$. For a small drift velocity [Fig.5(b)], the dynamics is not normal, however the asymptotic growth $G_{\infty}$ is relatively small ($G_{\infty} \simeq 1.2$), so that the system evolution does not deviate from the dominant bound state. As the drift velocity increases and approaches the critical one [Figs.5(c) and (d)], the  the asymptotic growth $G_{\infty}$ rapidly increases [$G_{\infty} \simeq 19$ and $G_{\infty} \simeq 366$ in Figs.5(c) and (d), respectively], small perturbations and noise in the system can be amplified, destroying the bound state, as shown in Figs.5(c) and (d). The figures clearly show that small perturbations are transiently amplified to power levels comparable or larger than the initial bound state mode, while being advected away from the potential well because of the drift. 
\section{Optical realization of anyonic $\mathcal{PT}$ symmetry} 
 The Schr\"odinger equation with Hamiltonian (5) can be of relevance to describe some optical or fluid systems, such as light dynamics in optical cavities or hydrodynamic flows \cite{r16,r18,r39bis,r25,r25tris}. In optics, the anyonic $\mathcal{PT}$ Hamiltonian (5) with non-vanishing phase $\varphi$ describes rather generally light dynamics in optical resonators and cavities with diffraction and spatial filtering, or temporal pulse propagation in the presence of material dispersion and finite gain bandwidth effects like in actively mode-locked laser systems \cite{r16,r18,r25}. For example, let us consider pulse dynamics in an actively mode-locked laser cavity containing an amplitude (AM) and a phase (FM) modulator, driven by the same periodic waveform $W(X)$ at frequency $\omega_m= 2 \pi /T_m$. The modulation frequency is set close to the separation $\Delta \omega_{ax}=2 \pi /T_R$ of cavity axial modes, where $T_R$ is the photon transit time in the resonator. The evolution of the pulse envelope $\psi(X,T)$ inside the optical cavity is described by the master equation \cite{r38,r39bis,ML1,ML2,ML3}
 \begin{equation}
 i T_R \frac{\partial \psi}{\partial T}=(-\mathcal{D}+i\mathcal{D}_g) \frac{\partial^2 \psi}{\partial X^2}+i(g-l)\psi+\Delta W(X-vT) \psi
 \end{equation}
 where $X$ is the fast time variable describing the pulse shape envelope in each cavity round trip ($-T_R/2<X<T_R/2$), $T$ is the slow time variable (the round-trip number) that describes pulse reshaping at successive transits in the cavity, $\mathcal{D}$ and $\mathcal{D}_g$ are the group velocity dispersion and spectral filtering parameters of the optical cavity, defined by the dispersive and gain elements of the resonator \cite{ML2,ML3}, $v=1-T_m/T_R$ is the normalized detuning parameter between cavity transit time and modulation period, $g$ and $l$ are the saturated gain and loss per round-trip in the resonator, $\Delta=\Delta_1-i\Delta_2$, and $\Delta_1$, $\Delta_2$ are the amplitudes of the FM and AM modulators, respectively.  Clearly, for $g=l$ and provided that the amplitudes $\Delta_1$ and $\Delta_2$ of modulators are tuned to satisfy the condition $\Delta_2/\Delta_1=\mathcal{D}_g / \mathcal{D}$, the master equation (31) defines a Hamiltonian dynamics with a non-Hermitian Hamiltonian $\mathcal{H}$ given by Eq.(5) where the anyonic phase $\varphi$ is given by 
 \begin{equation}
 \varphi={\rm atan}(\mathcal{D}_g/ \mathcal{D})
 \end{equation}
  and the potential is $V(X)  \propto W(X)$.  Note that the drift velocity $v$ can be tuned by changing the modulation frequency, with $v=0$ for exact synchronization $T_m=T_R$. Let us assume that $W(X)$ is real and describes a potential well, such as in the usual sinusoidal modulation. For exact synchronization $T_m=T_R$ of the modulation frequency, a stable mode-locked pulse, corresponding to the lowest bound state of the potential well, is established in the cavity, which saturates the gain and dominates the laser dynamics after an initial transient laser switch on.
For a sufficiently large detuning $v \neq 0$, corresponding to a modulation not synchronized with the photon transit time in the cavity \cite{ML1}, the disappearance of  bound states arising from non-Hermitian delocalization would  destroy stable mode-locking operation, which should be observable in an experiment.
 \section{Conclusion} 
In this Letter we introduced the concept of anyonic $\mathcal{PT}$-symmetry and disclosed intriguing scattering and localization properties that arise from the breakdown of the Galilean invariance of the Schr\"odinger equation, such as drift-induced transparency of a potential barrier and disappearance of bound states for fast moving potential wells. Such phenomena could be observed in photonic systems, such as in actively mode-locked lasers, where the optical pulse dynamics is effectively described by Hamiltonians with anyonic phase.


\begin{thebibliography}{0}


\bibitem{r1}
\Name{ Bender C.M. \and Boettcher S.}
\REVIEW{Phys. Rev. Lett.}{80}{1998}{5243}


\bibitem{r2}
\Name{ Bender C.M}
\REVIEW{Contemp. Phys.}{46}{2005}{277}

\bibitem{r3}
\Name{ Bender C.M}
\REVIEW{Rep. Prog. Phys.}{70}{2007}{947}

\bibitem{r4}
\Name{Konotop V.V., Yang J. \and Zezyulin D.A.}
\REVIEW{Rev. Mod. Phys.}{88}{2016}{035002}

\bibitem{r6}
\Name{Feng L., El-Ganainy R. \and Ge L.}
\REVIEW{Nat. Photon.}{11}{2017}{752}

\bibitem{r7} 
\Name{Longhi S.}
\REVIEW{EPL}{120}{2017}{64001}

\bibitem{r8}
\Name{Zhang Z., Ma D., Sheng J., Zhang Y., Zhang Y. \and Xiao M.}
\REVIEW{J. Phys. B}{51}{2018}{072001}

\bibitem{r12}
  \Name{Moiseyev N.}
  \Book{Non-Hermitian Quantum Mechanics}
  \Publ{Cambridge Univ.Press}
  \Year{2011}

\bibitem{r9}
\Name{Mostafazadeh A.}
\REVIEW{ J. Math. Phys. (N.Y.) }{43}{2002}{205}

\bibitem{r10}
\Name{Mostafazadeh A.}
\REVIEW{ J. Math. Phys. (N.Y.) }{43}{2002}{2814}

\bibitem{r11}
\Name{Mostafazadeh A.}
\REVIEW{Int. J. Geom. Methods Mod. Phys}{7}{2010}{1191}


\bibitem{r13}
\Name{Ge L. \and T\"ureci H.E.}
\REVIEW{Phys. Rev. A }{88}{2013}{053810}

\bibitem{r14}
\Name{Wu J.-H. , Artoni M. \and La Rocca G.C.}
\REVIEW{Phys. Rev. Lett.}{113}{2014}{123004}

\bibitem{r15}
\Name{Antonosyan D.A., Solntsev A.S. \and Sukhorukov A.A.}
\REVIEW{Opt. Lett.}{40}{2015}{4575}

\bibitem{r16}
\Name{Longhi S.}
\REVIEW{Ann. Phys.}{360}{2015}{150}


\bibitem{r17}
\Name{Peng P.,Cao W., Shen C.,Qu W., Wen J., Jiang L. \and Xiao Y.}
\REVIEW{Nat. Phys.}{12}{2016}{1139}


\bibitem{r18}
\Name{Longhi S.}
\REVIEW{Opt. Lett.}{41}{2016}{4518}


\bibitem{r19}
\Name{Yang F., Liu Y.-C. \and You L.}
\REVIEW{Phys. Rev. A}{96}{2017}{053845}

\bibitem{r19bis}
\Name{Kartashov Y.V., Konotop V.V. \and Zezyulin D.A.}
\REVIEW{EPL}{107}{2014}{50002}

\bibitem{refB}
\Name{Qi B., Zhang L. \and Ge L.}
\REVIEW{Phys. Rev. Lett.}{120}{2018}{ 093901}

\bibitem{long}
\Name{Longhi S.}
\REVIEW{Opt. Lett.}{43}{2018}{4025}

\bibitem{r20}
 \Name{Hatano N. \and Nelson D.R.}
\REVIEW{Phys. Rev. Lett.}{77}{1996}{570}


\bibitem{r21}
 \Name{Hatano N. \and Nelson D.R.}
\REVIEW{Phys. Rev. B}{58}{1998}{8384}


\bibitem{r22}
 \Name{Hatano N.}
\REVIEW{Physica A}{254}{1998}{317}


\bibitem{r23}
 \Name{Brouwer P.W., Silvestrov P.G. \and Beenakker C.W.J.}
\REVIEW{Phys. Rev. B}{56}{1997}{R4333 }

\bibitem{r24}
 \Name{Longhi S., Gatti D. \and Della Valle G.}
\REVIEW{Sci. Rep.}{5}{2015}{13376}

\bibitem{r26}
\Name{Rosen G.}
\REVIEW{Lett. Nuovo Cim.}{2}{1971}{61}

\bibitem{r27} 
\Name{Greenberger D.M.}
\REVIEW{Am. J. Phys.}{47}{1979}{35}

\bibitem{r27bis}
\Name{Muga J.G., Palao J.P., Navarro B. \and Egusquiza I.L.}
\REVIEW{Phys. Rep.}{395}{2004}{357}

\bibitem{r27tris}
\Name{Ruschhaupt A., Dowdall T., Simon M.A. \and Muga J.G.}
\REVIEW{EPL}{120}{2017}{20001}

\bibitem{r28}
\Name{Narevicius E., Serra P. \and Moiseyev N.}
\REVIEW{EPL}{62}{2003}{789}

\bibitem{r29}
 \Name{Longhi S., Gatti D. \and Della Valle G.}
\REVIEW{Phys. Rev. B}{92}{2015}{094204}

\bibitem{r30}
  \Name{Landau L.D. \and Lifshitz E.}
  \Book{Quantum Mechanics}
  \Publ{Pergamon, Oxford}
  \Year{1965}


\bibitem{r31}
 \Name{Longhi S.}
\REVIEW{EPL}{120}{2017}{20007}


\bibitem{r32}
 \Name{ Horsley S. A. R., Artoni M. \and La Rocca G.C.}
\REVIEW{Nat. Photon.}{9}{2015}{436}


\bibitem{r33}
 \Name{ Horsley S. A. R. \and Longhi S.}
\REVIEW{Am. J. Phys.}{85}{2017}{439}

\bibitem{r34}
\Name{Makris K.G., Ge L. \and T\"ureci H.E.}
\REVIEW{Phys. Rev. X}{4}{2014}{041044 }

\bibitem{r37}
\Name{Trefethen L.N., Trefethen A.E., Reddy S.C. \and Driscoll T.A.}
\REVIEW{Science}{261}{1993}{578}

\bibitem{r38}
\Name{K\"artner F.X., Zumb\"uhl D.M. \and Matuschek N.}
\REVIEW{Phys. Rev. Lett.}{82}{1999}{4428}

 \bibitem{r39}
\Name{Longhi S. \and Laporta P.}
\REVIEW{Phys. Rev. E}{61}{2000}{R989}







 
 
 \bibitem{r41}
\Name{Longhi S.}
\REVIEW{EPL}{115}{2016}{61001}


 

\bibitem{r39bis}
\Name{Longhi S.}
\REVIEW{Phys. Rev. E}{66}{2002}{056607}


\bibitem{r25}
\Name{Longhi S.}
\REVIEW{Phys. Rev. A}{90}{2014}{043827}


\bibitem{r25tris}
\Name{Reddy S.C. \and Henningson D.S.}
\REVIEW{J. Fluid Mech. }{252}{1993}{209}

\bibitem{ML1}
\Name{Longhi S.\and Laporta P.}
\REVIEW{Appl. Phys. Lett.}{73}{1998}{720}

\bibitem{ML2}
\Name{Haus H.A.}
\REVIEW{IEEE J. Sel. Top. Quantum Electron.}{6}{2000}{1173}

\bibitem{ML3}
\Name{Rana F., Ram R.J. \and Haus H.A.}
\REVIEW{IEEE J. Quantum Electron.}{40}{2004}{41} 


\end{thebibliography}
\end{document}